\documentclass{article}
\usepackage[T1]{fontenc}
\usepackage[latin9]{inputenc}
\usepackage{geometry}
\usepackage{xcolor}
\usepackage{array}
\usepackage{units}
\usepackage{textcomp}
\usepackage{multirow}
\usepackage{amsmath}
\usepackage{graphicx}



\begin{document}
\title{The $P\Phi$--Compromise Function as a criterion of merit
to optimize irreversible thermal engines}

\author{S. Levario--Medina\textsuperscript{a} and L. A. Arias--Hernandez\textsuperscript{b}\\
Departamento de F\'{i}sica, Escuela Superior de F\'{i}sica y Matem\'{a}ticas, Instituto\\
Polit\'{e}cnico Nacional, U. P. Zacatenco, edif. \#9, 2o Piso, Ciudad de M\'{e}xico,\\
07738, M\'{e}xico, levario@esfm.ipn.mx; http://orcid.org/0000-0003-4347-5068\textsuperscript{a} and\\ larias@esfm.ipn.mx; 
http://orcid.org/0000-0003-4054-5446; corresponding author\textsuperscript{b}}

\maketitle
\begin{abstract}
Several authors have proposed out of equilibrium thermal engines models,
allowing optimization processes involving a trade off between the
power output of the engine and its dissipation. These operating regimes
are achieved by using objective functions such as the ecological function
($EF$). In order to measure the quality of the balance between these
characteristic functions, it was proposed a relationship where power
output and dissipation are evaluated in the above mentioned $EF$--regime
and they are compared with respect to its values at the regime of
maximum power output. We called this relationship \textquotedbl Compromise
Function\textquotedbl{} and only depends of a parameter that measures
the quality of the compromise. Thereafter this function was used to
select a value of the mentioned parameter to obtain the generalization
of some different objective functions (generalizations of ecological
function, omega function and efficient power), by demanding that these
generalization parameters maximize the above mentioned functions. In this work we demonstrate that this function can be used directly as an objective function: the ``$P\Phi$--Compromise
Function'' ($C_{P\Phi}$), also that the operation modes corresponding
to the maximum Generalized Ecological Function, maximum Generalized
Omega Function and maximum Efficient power output, are special cases
of the operation mode of maximum $C_{P\Phi}$, having the same optimum
high reduced temperature, then the characteristic functions will be
the same in any of the above three working regimes, independent of
the algebraic complexity of each generalized function. These results
are presented for two different models of an irreversible energy converter:
a non--endoreversible and a totally irreversible, both with
heat leakage.
\end{abstract}

\section{Introduction}
In the last paragraph of the book \textquotedbl Reflections
on the motive power of heat and on machines fitted to develop that
power\textquotedbl{} S. Carnot writes\cite{carnot}:
\begin{quotation}
<<We should not expect ever to utilize in practice all the
motive power of combustibles. The attempts made to attain this result
would be far more hurtful than useful if they caused other important
considerations to be neglected. The economy of the combustible is
only one of the conditions to be fulfilled in heat--engines.
In many cases it is only secondary. It should often give up to safety,
to strength, to the durability of the engine, to the small space which
it must occupy, to small cost of installation, etc. To know how to
appreciate in each case, at their true value, the considerations of
convenience and economy which may present themselves; to know how
to discern the more important of those which are only accessories;
to balance them properly against each other, in order to attain the
best results by the simplest means; such should be the leading characteristics
of the man called to direct, to co-ordinate among themselves the labors
of his comrades, to make them cooperate towards one useful end, of
whatsoever sort it may be.>>
\end{quotation}
In modern language these words give rise to the process known
as Thermodynamic Optimization. Since 1975, when Curzon and Ahlborn
proposed the power output of their model for thermal engines \cite{Curzon-Ahlborn},
as an objective function to find a specific operating regime, a large
number of articles have been published with several objective functions
which are associated with modes of operation or specific designs \cite{reviews,chen04}.
Thus, the studies of the behavior of engines have focused on finding
``optimal'' operating regimes, by means of economic, ecological
or other reasons. To obtain a model that allows to know the conditions
with which it is possible to reach such working regimes, given particular
operational objectives, in the framework of the Finite Time Thermodynamics
(FTT) have emerged objective functions, such as ``Ecological function'',
``Omega function'', ``Efficient Power'' among others \cite{FEcologica,FunOmg,stucki80,yilmaz06,arias09,calvo00},
that allow to find modes of operation that satisfy some compromise
between the characteristic functions of the engine.

The ecological functions is defined as \cite{FEcologica},
\begin{equation}
E=P-\Phi,\label{eq:FunEco}
\end{equation}
where $P$ is the power output and $\Phi$ the dissipation
of the engine. This function was proposed by Angulo-Brown towards
1992. However, the necessity to find the best objective function that
made a good trade off between the mentioned process variables, led
to the generalization of the ecological function in 1997 \cite{FEcoGen},
\begin{equation}
E_{G}=P-\epsilon\Phi,\label{eq:FunEcoGen}
\end{equation}
being $\epsilon$ a parameter that generate a family of ecological
functions ($E_{G}$). To choice the ecological
function that gives the best trade off between $P$ and $\Phi$, the
``Compromise Function'' $C$ \cite{FEcoGen,FunCom2,FunCom3} was
proposed as follows:
\begin{equation}
C=\frac{P^{MX}}{P^{MP}}-\frac{\varPhi^{MX}}{\varPhi^{MP}},\label{eq:DefFunCompromiso}
\end{equation}
where the super index $MX$ indicates that the power output and dissipation
must be evaluated in the regime of maximum $E_{G}$ ($ME_{G}$) and
$MP$ super index of power output and dissipation means that this
characteritic functions must be evaluated at the maximum power output
($MPO$) regime. Under this consideration, the ecological function
provides $75\%$ of the power output and $25\%$ of the dissipation
with respect to the $MPO$--regime for an endoreversible
model ($75-25$ corollary \cite{FEcoGen}). In 2001, Calvo et al \cite{FunOmg}
defined an optimiztion criterion, called the Omega
criterion, consisting in the maximization of the function 
\begin{equation}
\Omega=E_{u,eff}-E_{u,l},\label{omega}
\end{equation}
where $E_{u,eff}\equiv E_{u}-z_{min}E_{i}$ is the effective useful
energy and $E_{u,l}\equiv z_{max}E_{i}-E_{u}$ is the lost useful
energy. This functions were defined in terms of: $E_{u}$the useful
energy, $E_{i}$ the input energy and $z=E_{u}/E_{i}$ the mesure
of the engine's performance. In 2006 Partido and Arias-Hernández found
that the generalization of Omega function \cite{OmegaGen}: 
\begin{equation}
\Omega_{G}=E_{u,eff}-\lambda E_{u,l},\label{eq:DefFunOmg}
\end{equation}
could be equivalent to the \textquotedbl best\textquotedbl{} of the
generalized ecological ones. This happens when the $\lambda$ parameter
is selected through the Compromise Function. In 2016 Levario-Medina
and Arias-Hernandez found the same equivalence by picking the parameter
$k$ of the $k$--Efficient Power \cite{kPotEfi,levario19}:
\begin{equation}
P\eta_{k}=P\eta^{k},\label{eq:DefkPotEfi}
\end{equation}
with the same procedure by using the Compromise Function. In reference
\cite{kPotEfi} was showed that the above procedure to select the
generalization parameter, allow us to make equivalent the optimal
regimes derived from the three mentioned objective functions ($E_{G}$,
$\Omega_{G}$ and $P\eta_{k}$), in spite of they involve different
process variables and have different algebraic structure.

In this paper we show that the Compromise Function $C_{P\Phi}$ defined
as:
\begin{equation}
C_{P\varPhi}\left(a_{h}\right)=\frac{P\left(a_{h}\right)}{P^{MP}}-\frac{\Phi\left(a_{h}\right)}{\varPhi^{MP}}.\label{eq:DefFunCPPhiFundeCom}
\end{equation}
can be taken directly as an objective function without the use of
the generalized functions. This objective function that from now on
we will call $P\Phi$--Compromise Function, allow us to
reach the best trade--off between $P$ and $\Phi$ of the
thermal engine ($75-25$ corollary) \cite{arias09,FEcoGen,FunCom2}.

The present study is based on an akin model to the one proposed by
Curzon and Ahlborn in 1975 (CA--engine). A CA--engine
Fig. \ref{fig:MotorTipoCAN} has two reservoirs at absolute temperatures
$T_{1}$ and $T_{2}$ such that $T_{1}>T_{2}$;
two irreversible components (thermal conductances $\alpha$
and $\beta$) and a working substance that operates in reversible
cycles between two working temperatures $T_{1w}$ and
$T_{2w}$ ($T_{1w}$ > $T_{2w}$) \cite{Curzon-Ahlborn}. However,
this model still needs certain elements that have been added later
by other authors, elements which consider internal irreversibilities
($\sigma_{i}$) \cite{Tollman,Gordon2,ParaRNendo,Arias2013} or the
heat leaks that occur through the materials with which the thermal
engines are built. To emulate the heat leaks, a direct heat flux is
considered between the reservoirs ($Q_{hl}$). Additionally, we have
two heat fluxes $Q_{1}$ and $Q_{2}$ (quantities per cycle period),
which allow us to build the other processes variables of the engine:
\begin{itemize}
\item The power output,
\begin{equation}
P=Q_{I}-Q_{O}.\label{pot}
\end{equation}
\item The efficiency,
\begin{equation}
\eta=1-\frac{Q_{O}}{Q_{I}}.\label{ef}
\end{equation}
\item The entropy production,
\begin{equation}
\sigma_{T}=\left(\frac{Q_{O}}{T_{2}}-\frac{Q_{I}}{T_{1}}\right)+\left(\frac{Q_{1}}{T_{1w}}-\frac{Q_{2}}{T_{2w}}+\sigma_{i}\right)\geq0.\label{ep}
\end{equation}
being the first parenthesis the entropy production of the surroundings
($\sigma_{e}$), the second one can be related with the entropy production
of the internal cycle ($\sigma_{ws}$), due to the working substance
operates in cycles $\sigma_{ws}$ always is zero, that is, $\sigma_{ws}=0$;
$\sigma_{i}$ is the entropy production which arise in the working
substance, due to different processes such as turbulence and
viscosity among others secondary processes which can help to complete
the cycle. On the other hand, $Q_{I}$ is the total input heat flux to the system, and $Q_{O}$ its the total output heat flux. In general,
these heat fluxes are:
\begin{equation}
Q_{I}=Q_{1}+Q_{hl}\label{eq:Qi}
\end{equation}
and 
\begin{equation}
Q_{O}=Q_{2}+Q_{hl}.\label{eq:Qc}
\end{equation}
In this work, we consider a linear heat transfer law (Newtonian heat
law) for all heat fluxes involved (see Fig. \ref{fig:MotorTipoCAN}),
\begin{figure}[t]
\begin{centering}
\includegraphics[width=14cm, height=7cm]{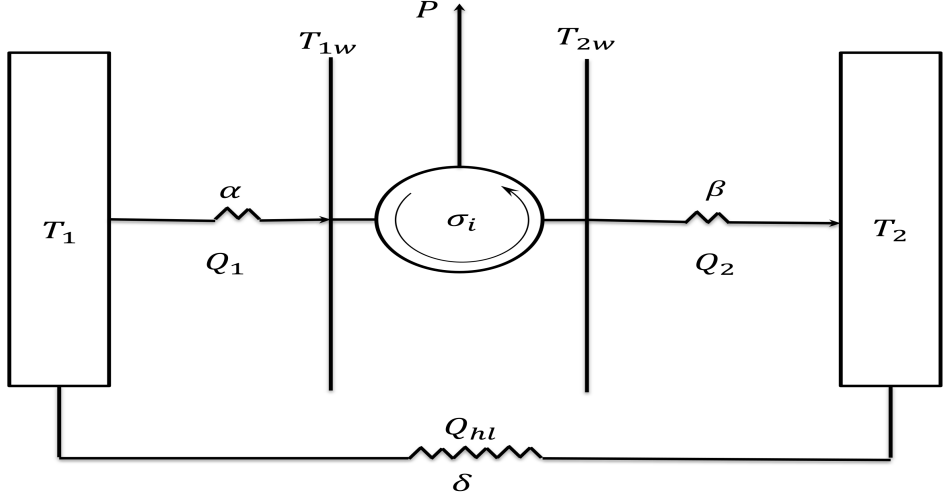}
\par\end{centering}
\caption{\label{fig:MotorTipoCAN} Heat engine akin Curzon--Ahlborn
thermal engine.}
\end{figure}
that is, 
\begin{equation}
Q_{1}=\alpha\left(T_{1}-T_{1w}\right),\label{eq:Q1}
\end{equation}
\begin{equation}
Q_{2}=\beta\left(T_{2w}-T_{2}\right)\label{eq:Q2}
\end{equation}
and
\begin{equation}
Q_{hl}=\delta\left(T_{1}-T_{2}\right).\label{eq:Qhl}
\end{equation}
For the models here used, we also build a characteristic function
called the dissipation function, defined as \cite{FEcologica},
\begin{equation}
\Phi=T_{2}\sigma_{T}.\label{eq:dis}
\end{equation}
\end{itemize}
The equations (\ref{eq:Q1}), (\ref{eq:Q2}) and (\ref{eq:Qhl}),
can be rewritten as:
\begin{equation}
Q_{1}\left(\alpha,T_{1},a_{h}\right)=\alpha T_{1}\left(1-a_{h}\right),\label{eq:Q1FdP}
\end{equation}

\begin{equation}
Q_{2}\left(\alpha,\gamma,T_{1},a_{c}\right)=\alpha T_{1}\frac{\tau}{\gamma}\left(\frac{1}{a_{c}}-1\right)\label{eq:Q2FdP}
\end{equation}
and
\begin{equation}
Q_{hl}\left(\delta,\tau,T_{1}\right)=\delta T_{1}\left(1-\tau\right),\label{eq:QhlFdP}
\end{equation}
where $\tau=T_{2}/T_{1}$ and $\gamma=\alpha/\beta$, while $a_{h}=T_{1w}/T_{1}$
and $a_{c}=T_{2}/T_{2w}$ are the high reduced temperature and the
low reduced temperature respectively. These last variables will be
important, because they help us to characterize several modes of operation.
Since the thermal engine operates in cycles, we could find a relation
between these reduced temperatures, depending of the model which will
be used.

Hereinafter there are certain considerations must be made to obtain
each of the models used in this work, for each of them we will show
that the $P\Phi$--Compromise Function is an objective function,
that allow us characterizing an \textquotedbl optimal\textquotedbl{}
mode of operation whose properties were attributed to the generalized
ecological function and others objective functions. We made this analysis
for two irreversible models. In section 2, a non--endoreversible
model with heat leak (NEHL) is addressed. In this model the irreversibilities
of the internal cycle are quantified by the non-endoreversibility
parameter $R$ \cite{ParaRNendo} and a heat leak is added between
the two heat reservoirs through a thermal conductance, this last with
the purpose of obtaining a loop shaped characteristic curve of the
power output versus efficiency, reported by Gordon in the eighties
for real thermal engines\cite{Gordon2,Gordon1,Gordon3}. In section
3, the ``uncompensated heat'' of Clausius \cite{CalNcomClas} is
used as a measure of the irreversibilities that are generated in the
working substance, likewise a \textquotedbl heat leak\textquotedbl{}
is incorporated with the purpose of obtaining an Irreversible model
with Heat Leak (IHL), in which the most important sources of irreversibilities
that occur in a real energy converter are included. In addition, in
each of these sections we show how, under the appropriate considerations,
the results corresponding to an endoreversible model can be obtained.
Finally, an appendix is added to show how the compromise function
is used to select the parameter of generalization of any of the objective
functions above mentioned, to obtain the same optimal operation regime.

\section{Non--endorreversible model with heat leak {[}$\sigma_{i}=(1-R)$;
$\delta\protect\neq0${]}}
In the CA model, Curzon and Ahlborn supposed that the work substance
operates in such way, that its internal entropy production ($\sigma_{i}$)
was zero, what is a great supposition, because in nature, these akin
of processes are not common, if not nonexistent. Due to this, some
authors have proposed to add a phenomenological parameter ($R$) \cite{ParaRNendo},
which allow us to quantify the grade of irreversibility that is generated
within the work substance. So that, the $R$ parameter converts the
Clausius's inequality \cite{Zemansky} to an equality. It permits
to obtain a better approximation to a real thermal engines behavior.
Under this consideration the entropy production of the working substance
$\left(\sigma_{i}\right)$ can be written as:
\begin{equation}
\sigma_{i}=\left(1-R\right)\frac{Q_{2}}{T_{2w}},\label{eq:EntroCicloR}
\end{equation}
with $0<R\leq1$, (when $R=1$, the endoreversible mode ($\sigma_{i}=0$)
is recuperated). Replacing the equation (\ref{eq:EntroCicloR}) in
the second parentheses of the equation (\ref{ep}), $\sigma_{ws}$
can be rewritten as:
\begin{equation}
\frac{Q_{1}}{T_{1w}}-R\frac{Q_{2}}{T_{2w}}=0,\label{eq:HNendoR}
\end{equation}
this one is know as the non-endoreversibility hypothesis. From this
relationship and considering the heat flows $Q_{1}$ and $Q_{2}$
given by the equations (\ref{eq:Q1}) and (\ref{eq:Q2}) respectively,
it is possible to establish the above mentioned relation between the
high reduced temperature and the low reduced temperature:
\begin{equation}
a_{c}\left(\gamma,R,a_{h}\right)=1+\frac{\left(a_{h}-1\right)\gamma}{a_{h}R}.\label{eq:acNendoR}
\end{equation}
This allows to written the heat flux $Q_{2}$ like:
\begin{equation}
Q_{2}\left(\alpha,\gamma,T_{1},\tau,R,a_{h}\right)=\frac{\left(1-a_{h}\right)T_{1}\alpha\tau}{a_{h}\left(R+\gamma\right)-\gamma}\label{eq:QcNendoR}
\end{equation}
and $Q_{1}$is given by the equation (\ref{eq:Q1FdP}).

Then the heat fluxes which are get in ($Q_{I}=Q_{1}+Q_{hl}$) and
get out ($Q_{O}=Q_{2}+Q_{hl}$) of the system are:
\begin{equation}
Q_{I}\left(\alpha,\delta,\tau,T_{1},a_{h}\right)=T_{1}\left[\alpha\left(1-a_{h}\right)+\delta\left(1-\tau\right)\right]\label{eq:QEnEndoCCR}
\end{equation}
and
\begin{equation}
Q_{O}\left(\alpha,\delta,\gamma,\tau,T_{1},a_{h}\right)=\frac{T_{1}\left\{ a_{h}\left(R+\gamma\right)\delta+\gamma\delta\left(\tau-1\right)+\alpha\tau-a_{h}\left[\alpha+\left(R+\gamma\right)\delta\right]\tau\right\} }{a_{h}\left(R+\gamma\right)-\gamma}.\label{eq:QSnEndoCCR}
\end{equation}
Thus, the power output and the dissipation of the system will be: 
\begin{equation}
P\left(\alpha,\gamma,T_{1},\tau,R,a_{h}\right)=T_{1}\alpha\left(a_{h}-1\right)\left[\frac{\tau}{a_{h}\left(R+\gamma\right)-\gamma}-1\right]\label{eq:PotNendoCCR}
\end{equation}
and
\begin{equation}
\Phi\left(\alpha,\delta,\gamma,T_{1},\tau,R,a_{h}\right)=\frac{1}{a_{h}\left(R+\gamma\right)-\gamma}\left\{ \begin{array}{c}
\alpha\tau+a_{h}^{2}\alpha\left(R+\gamma\right)\tau-\gamma\left[\delta\left(\tau-1\right)^{2}-\alpha\tau\right]\\
+a_{h}\left\{ \left(R+\gamma\right)\delta-\left[\alpha\left(1+R+2\gamma\right)+2\left(R+\gamma\right)\delta\right]\tau+\left(R+\gamma\right)\delta\tau^{2}\right\} 
\end{array}\right\},\label{eq:DisMNoEndoRCC}
\end{equation}
The power output is a convex function, with its maximum in:
\begin{equation}
a_{h}^{MP}\left(\gamma,\tau,R\right)=\frac{\gamma+\sqrt{R\tau}}{R+\tau}.\label{eq:ahMPnEndoR}
\end{equation}

Whereupon, and substituting the definitions (\ref{eq:PotNendoCCR}),
(\ref{eq:DisMNoEndoRCC}) and (\ref{eq:ahMPnEndoR}) in (\ref{eq:DefFunCPPhiFundeCom}),
the function $C_{P\Phi}$ will be:
\begin{equation}
C_{P\Phi}\left(\alpha,\delta,\gamma,\tau,R,a_{h}\right)=\frac{R\kappa\tau\left(\tau-1\right)\left\{ a_{h}^{2}\kappa\left[\kappa\delta\left(\tau-1\right)+\alpha\left(\tau-\sqrt{R\tau}\right)\right]+a_{h}C_{n1}+C_{n0}\right\} }{\left(a_{h}\kappa-1\right)\left(\sqrt{R\tau}-\tau\right)\left\{ \sqrt{R\tau}\left[\gamma\delta\left(\tau-1\right)^{2}-\alpha\tau\right]+R\left[\delta\sqrt{R\tau}\left(\tau-1\right)^{2}+\alpha\tau\left(1+\tau-\sqrt{R\tau}\right)\right]\right\} },\label{eq:CpPhiNendoRCC}
\end{equation}
with
\begin{equation}
C_{n1}\left(\alpha,\delta,\gamma,T_{1},\tau,R,a_{h}\right)=.\left[\alpha\sqrt{R\tau}-2\delta\left(\tau-1\right)\left(\gamma+\sqrt{R\tau}\right)\right]+2\gamma\left[-\delta\sqrt{R\tau}\left(\tau-1\right)+\alpha\left(\sqrt{R\tau}-\tau\right)\right]-2\gamma^{2}\delta\left(\tau-1\right)-\alpha\tau\sqrt{R\tau}\label{eq:Cn1NendoRCC}
\end{equation}
and
\begin{equation}
C_{n0}\left(\alpha,\delta,\gamma,T_{1},\tau,R,a_{h}\right)=\gamma^{2}\delta\left(\tau-1\right)+\tau\left[R\delta\tau+\alpha\sqrt{R\tau}-R\left(\alpha+\delta\right)\right]+\gamma\left[2\delta\sqrt{R\tau}\left(\tau-1\right)+\alpha\left(\tau-\sqrt{R\tau}\right)\right]\label{eq:Cn0NendoRCC}
\end{equation}
\begin{figure}[t]
\begin{centering}
\includegraphics[width=15cm, height=6cm]{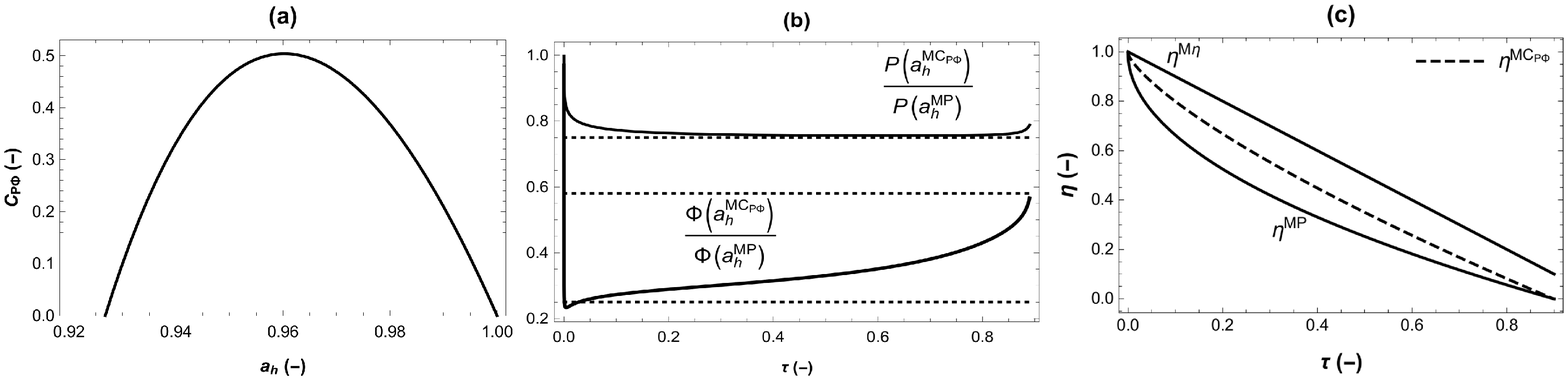}
\par\end{centering}
\caption{\label{fig:FunComNendoRCC}(a) $P\Phi$--Compromise function
under a non--endoreversible model as a function of the high
reduced temperature, (b) Comparison of the power output and dissipation
at maximum $P\Phi$\textendash{} Compromise Function with respect
to the same process variables at maximum power output regime and (c)
Efficiency at maximum efficiency regime ( $\eta^{M\eta}$), efficiency
at maximum power output ($\eta^{MP}$) and efficiency at maximum $P\Phi$
\textendash{} Compromise Function (Dashed line, $\eta^{MC_{P\Phi}}$).
Here we consider: $\alpha=1\,MW/K$, $\gamma=3$,$\tau=0.5$, $T_{1}=500K$,
$\delta=0.001MW/K$ and $R=0.9$.}
\end{figure}

In this function, as can be observed in Fig \ref{fig:FunComNendoRCC}
(a), the high reduced temperature that allows us to reach its maximum
is:
\begin{equation}
a_{h}^{MC_{P\Phi}}\left(\alpha,\delta,\gamma,T_{1},\tau,R,a_{h}\right)=\frac{\gamma}{\left(R+\gamma\right)}-\frac{r_{Cp}}{\left(R+\gamma\right)\left[\alpha\left(\tau-\sqrt{R\tau}\right)-\kappa\delta\left(1-\tau\right)\right]},\label{eq:ahMCpPhiNEndoRCC}
\end{equation}
where
\begin{equation}
r_{Cp}\left(\alpha,\delta,\gamma,T_{1},\tau,R,a_{h}\right)=\sqrt{R\tau\left[\alpha\sqrt{R\tau}-\gamma\delta\left(1-\tau\right)-R\left(\alpha+\delta\left[1-\tau\right]\right)\right]\left[\alpha\left(\tau-\sqrt{R\tau}\right)-\kappa\delta\left(1-\tau\right)\right]}.\label{eq:rCpNendoRCC}
\end{equation}

This high reduced temperature characterize a particular energetic
performance in the engine's behavior. Its process variables can be
obtained substituting the equations (\ref{eq:QEnEndoCCR}) and (\ref{eq:QSnEndoCCR}),
with the $a_{h}$ given by equation (\ref{eq:ahMCpPhiNEndoRCC}),
as is indicated in equations (\ref{pot}), (\ref{ef}) and (\ref{eq:dis}).
This allows the engine to reach a 75\% of the power output and around
25\% to 60\% of the dissipation of the MPO--regime (as is
showed in Fig. \ref{fig:FunComNendoRCC} (b)). In Chapter 3 section
3.2.3 of reference \cite{kPotEfi} (also see appendix A) is showed
how to obtain the high reduced temperature that characterize the $ME_{G}$--regime,
using the Compromise Function like is defined by the equation (\ref{eq:DefFunCompromiso}).
By substituting equation (\ref{eq:EpsMaxFunCompPotDisNendoRCC}) in
(\ref{eq:ahMaxEfiMNoEndoRCC}) (see the appendix A) it is possible
to observe that the $a_{h}$ which characterize the maximum generalized
ecological function regime, is the same that is given by equation
(\ref{eq:ahMCpPhiNEndoRCC}). So that, the energetic performance of
a thermal engine working in any of these regimes are going to be equal.
Under the appropriate limit conditions ($\delta\rightarrow0$ y $R\rightarrow1$),
this result can be reduced to the case of an endorreversible model,
as is showed in Table \ref{tab:RrNEHLtoE}.
\begin{table}[tb]
\begin{centering}
\begin{tabular}{|c|>{\centering}m{10cm}|c|}
\hline 
$C_{P\Phi}$ & $-\frac{\left[a_{h}-1\right]\left[1+\gamma\right]\left[a_{h}+\left(a_{h}-1\right)\gamma-\sqrt{\tau}\right]\left[1+\sqrt{\tau}\right]}{\left[a_{h}+\left(a_{h}-1\right)\gamma\right]\left[-1+\sqrt{\tau}\right]^{2}}$ & Sketch of the compromise function\tabularnewline
\cline{1-2} \cline{2-2} 
$a_{h}^{MC_{P\Phi}}$ & $\frac{\gamma+\tau^{\nicefrac{1}{4}}}{1+\gamma}$ & \multirow{3}{*}{\includegraphics[scale=0.6]{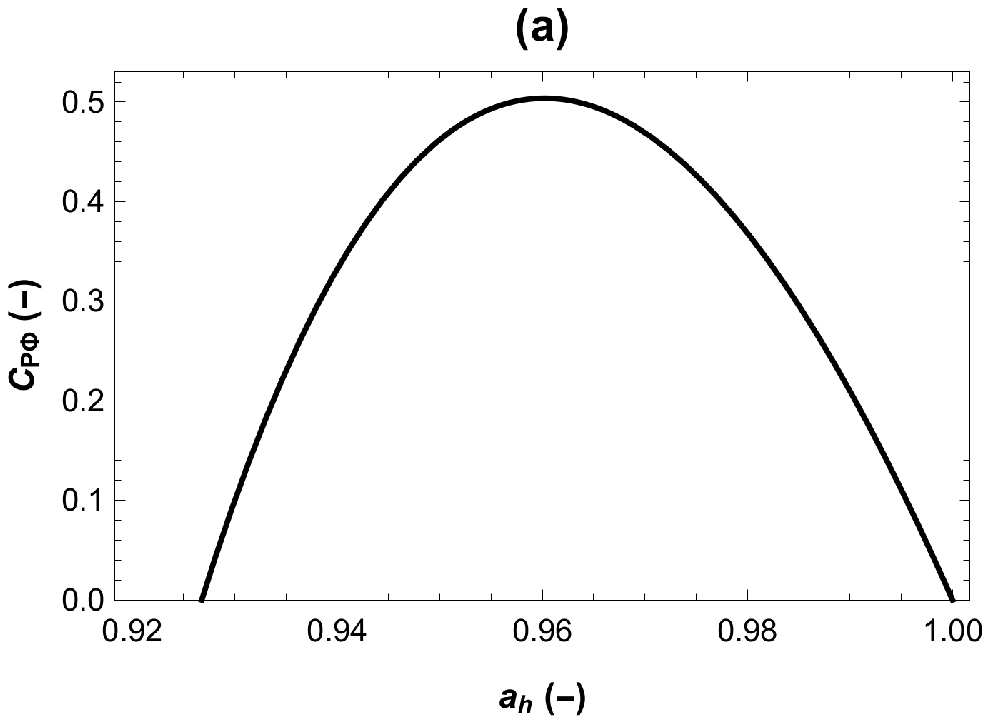}}\tabularnewline
\cline{1-2} \cline{2-2} 
\multicolumn{2}{|c|}{Comparative} & \tabularnewline
\multicolumn{2}{|c|}{\includegraphics[scale=0.5]{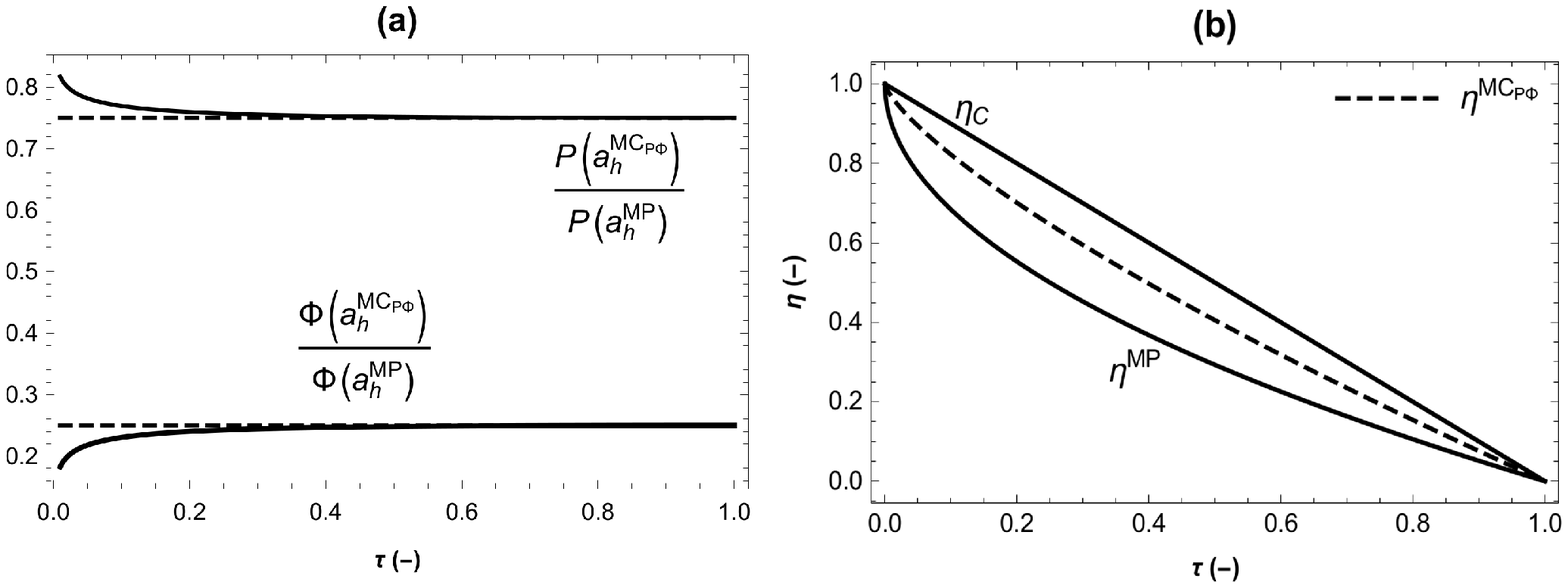}} & \tabularnewline
\hline 
\end{tabular}
\par\end{centering}
\caption{\label{tab:RrNEHLtoE}Results reduction to an endoreversible model
without heat leak.}
\end{table}

\section{Irreversible model with heat leak ($\sigma_{i}\protect\geq0$; $\delta\protect\neq0$)}
As is mentioned in the previous section, the non-endorreversibility
parameter is an option to measure possible irreversibilities which
could arise in the working substance of the internal cycle. However,
the parameter $\sigma_{i}$ in the second parenthesis of the equation
(\ref{ep}) can be associated directly to the Uncompensated Heat of
Clausius (UHC). This concept was proposed by Clausius \cite{CalNcomClas,FWtC},
to offset the heat which is lost in a irreversible process. Under
this consideration, the second parenthesis of the
equation (\ref{ep}) allows to establish the relationship between
the high reduced and the low reduced temperatures which is:
\begin{equation}
a_{c}\left(\alpha,\gamma,r,a_{h}\right)=1+\gamma\left(1-r\right)-\frac{\gamma}{a_{h}},\label{eq:acfahCnCC}
\end{equation}
where $r=\nicefrac{\sigma_{i}}{\alpha}$. So, the heat flux $Q_{2}$
can be rewritten as:
\begin{equation}
Q_{2}\left(\alpha,\gamma,T_{1},\tau,r,a_{h}\right)=\frac{\alpha T_{1}\tau\gamma\left[a_{h}\left(a_{h}-r\right)-1\right]}{\gamma+a_{h}\left[\gamma\left(r-1\right)-1\right]}\label{eq:Q2TI}
\end{equation}
and $Q_{1}$ is given by the equation (\ref{eq:Q1FdP}).

Even though, the characteristic loop of thermal engines out of equilibrium
between the efficiency and the power output could be generated by
using only the UHC, which implies that this is a good way to measure
the irreversibilities of the internal cycle, to incorporate the most
common sources of irreversibility, in this model, a heat leak is added.
This element will operate like in the previous section, with a newtonian
law given by the equation (\ref{eq:Qhl}). So, the heat fluxes which
are get into and get out of the system are given by the equations
(\ref{eq:Qi}) and (\ref{eq:Qc}):
\begin{equation}
Q_{IC}\left(\alpha,\delta,T_{1},\tau,a_{h}\right)=T_{1}\left[\alpha\left(1-a_{h}\right)+\delta\left(1-\tau\right)\right]\label{eq:QE4C}
\end{equation}
and
\begin{equation}
Q_{OC}\left(\alpha,\delta,\gamma,T_{1},\tau,r,a_{h}\right)=T_{1}\frac{\gamma\delta\left(\tau-1\right)+\alpha\tau+a_{h}\left\{ \delta\left(\tau-1\right)\left[\gamma\left(r-1\right)-1\right]+\alpha\tau\left(r-1\right)\right\} }{a_{h}\left[1-\gamma\left(r-1\right)\right]-\gamma},\label{eq:QS4C}
\end{equation}
with these equations, it is possible to establish the power output
and the dissipation of this model as:
\begin{equation}
P\left(\alpha,\delta,\gamma,T_{1},\tau,r,a_{h}\right)=T_{1}\alpha\left\{ 1-a_{h}+\frac{\tau\left[1+a_{h}\left(r-1\right)\right]}{\gamma+a_{h}\left[\gamma\left(r-1\right)-1\right]}\right\} \label{eq:Pot4C}
\end{equation}
and
\begin{equation}
\Phi\left(\alpha,\delta,\tau,T_{1},r,a_{h}\right)=T_{1}\frac{a_{h}^{2}\alpha\tau\left[\gamma\left(r-1\right)-1\right]-\alpha\tau+\gamma\left[\delta\left(1-\tau\right)^{2}-\alpha\tau\right]+a_{h}\left\{ \delta\left(1-\tau\right)^{2}\left[\gamma\left(r-1\right)-1\right]-\alpha\tau\left(1+\gamma\right)\left(r-2\right)\right\} }{\gamma+a_{h}\left[\gamma\left(r-1\right)-1\right]}.\label{eq:Dis4C}
\end{equation}
Here, power output has a high reduced temperature that allow its maximization.
This $a_{h}$is:
\begin{equation}
a_{h}^{MP}\left(\gamma,\tau,r\right)=\frac{\gamma+\sqrt{\tau}}{1-\gamma\left(r-1\right)}.\label{eq:ahMaxPot4C}
\end{equation}
As was mentioned in previous section, when equations (\ref{eq:Pot4C}),
(\ref{eq:Dis4C}) and (\ref{eq:ahmCpPhi4C}) are substituted in (\ref{eq:DefFunCPPhiFundeCom}),
the objective function $C_{P\Phi}$ is given by:
\begin{equation}
C_{P\Phi}\left(\alpha,\delta,\gamma,\tau,r,a_{h}\right)=\nicefrac{\left\{ \begin{array}{c}
\left(\tau-1\right)\left[\gamma\left(r-1\right)-1\right]\\
\times\left\{ \gamma+a_{h}\left[\gamma\left(r-1\right)-1\right]+\sqrt{\tau}\right\} n_{C1}
\end{array}\right\} }{\left\{ \begin{array}{c}
\left\{ \gamma+a_{h}\left[\gamma\left(r-1\right)-1\right]\right\} \left[r\left(\gamma+\tau\right)-\left(\sqrt{\tau}-1\right)^{2}\right]\\
\times\left\{ \begin{array}{c}
\delta\left(\tau-1\right)^{2}\left[\gamma\left(r-1\right)-1\right]\\
-\sqrt{\tau}\alpha\left\{ 1+\left[r\left(1+\gamma\right)-2\right]+\tau\right\} 
\end{array}\right\} 
\end{array}\right\} },\label{eq:CpPhi4C}
\end{equation}
with 
\begin{equation}
n_{C1}\left(\alpha,\delta,\gamma,\tau,r,a_{h}\right)=\left\{ \begin{array}{c}
\gamma\delta\left(\tau-1\right)+\left[\alpha\sqrt{\tau}+\delta\tau-\left(\alpha+\delta\right)\right]\sqrt{\tau}\\
+a_{h}\left\{ \delta\left(\tau-1\right)\left[\gamma\left(r-1\right)-1\right]+\alpha\sqrt{\tau}+\alpha\tau\left(r-1\right)\right\} 
\end{array}\right\} .\label{eq:nc14C}
\end{equation}

Here, the high reduced temperature that maximizes $C_{p\Phi}$ is (see Fig. 3(a)):
\begin{figure}[t]
\begin{centering}
\includegraphics[width=15cm, height=10cm]{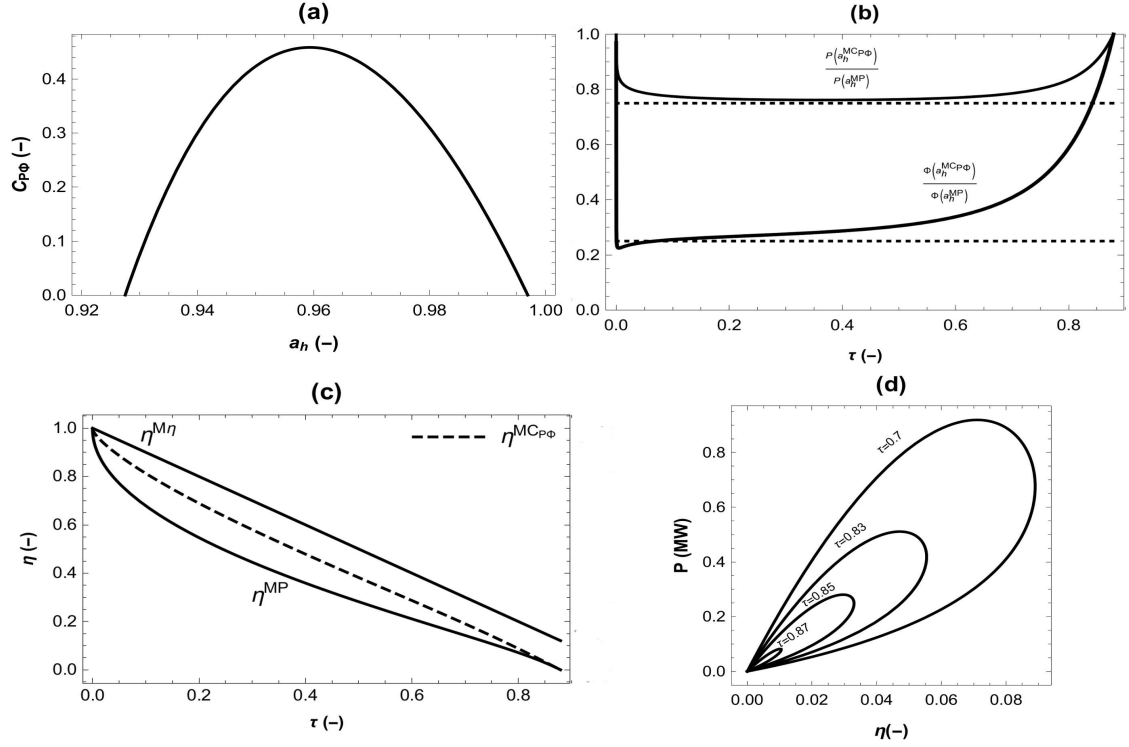}
\par\end{centering}
\caption{\label{fig:FunComCnCC}(a) $P\Phi$ \textendash{} Compromise Function
for an irreversible model with heat leak. (b) Comparison of the power
output and dissipation at maximum compromise function with respect
to the same process variables at maximum power output regime. (c)
Efficiency at maximum efficiency regime ($\eta^{M\eta}$), efficiency
at maximum power output regime ($\eta^{MP}$) and efficiency at \textit{MCF}--regime
(dashed line, $\eta^{MC_{P\Phi}}$). (d) Parametric curves of power
output vs efficiency at different values of $\tau$. Here we use:
$\alpha=1\,MW/K$, $\gamma=3$, $\tau=0.5$, $T_{1}=500K$, $\delta=0.001MW/K$
and $r=0.001$.}
\end{figure}
\begin{equation}
a_{h}^{MC_{P\Phi}}\left(\alpha,\delta,\gamma,\tau,r\right)=\left[\frac{1}{1-\gamma\left(r-1\right)}\right]\left\{ \gamma+\sqrt{\frac{\tau\left\{ \delta\left[\tau-1\right]\left[\gamma\left(r-1\right)-1\right]-\alpha\left[r\gamma+\sqrt{\tau}-1\right]\right\} }{\delta\left[\tau-1\right]\left[\gamma\left(r-1\right)-1\right]+\alpha\left[\sqrt{\tau}+\tau\left(r-1\right)\right]}}\right\} .\label{eq:ahmCpPhi4C}
\end{equation}
In Fig. \ref{fig:FunComCnCC} (b) we can observe the comparative
of the power output and the dissipation at Maximum $P\Phi$--Compromise
Function ($MCF$--regime) with respect to the $MPO$--regime
characteristic functions. There, the quotient of the power output
at $MCF$--regime and power output at $MPO$--regime
is around 75\% until it is near to:
\[
\tau_{l}\left(\gamma,r\right)=\frac{\left[1-\sqrt{r\left[1+\gamma\left(1-r\right)\right]}\right]^{2}}{\left[1-r\right]^{2}},
\]
while dissipation increases considerably until 1, as $\tau$ approaches
to $\tau_{l}$. It implies that the energetic will be the same of
the MPO--regime. It has sense because $\tau$ is $T_{2}/T_{1}$
and it is relate with the engine's thermal gradient. When $\tau$
is near to zero ($T_{2}\ll T_{1}$), there are more configurations
that allow the thermal engine access to several modes of operation.
This fact can be observed in Fig. \ref{fig:FunComCnCC} (d), where
the parametric loop power output vs efficiency decreases as $\tau$
increases. If the reservoirs were bodies with finite heat capacities,
they can transfer more energy before reaching the thermal equilibrium
when $\tau\ll1$, in the other hand (if $\tau\approx1$) the quantity
of energy transferred is small before the thermal equilibrium can
be reached.

As in the above section, for this model there are conditions which
get it equivalent with the endoreversible model, they are $r=0$ and
$\delta=0$. With these conditions we obtain the same results showed
in Table \ref{tab:RrNEHLtoE}. However, no conditions exist which
allow the equivalence to this model with the NEHL model.

\section{Conclusions}
In this paper has been shown how by means of a linear heat transfer
law and two different thermal engine models (in which some important
features of the real thermal engines have been incorporated), the
$P\Phi$--Compromise Function is capable to lead a thermal
engine to reach a particular optimal operation mode, even in the endoreversible
limit ($R\rightarrow1$ or $r\rightarrow0$ and $\delta\rightarrow0$).
Such mode had been associated with the maximum ecological generalized
function regime, providing a 75\% of the power output and a 25\% of
the dissipation, compared to the same characteristic functions of
the MPO--regime. Despite of the fact that the generalized
ecological function was the first objective function which allows
to characterize the above mentioned optimal operation mode, through
a second process of optimization in which the Compromise Function
is used to select a very particular value of its generalization parameter,
in resent years, there are at least two more generalizations (generalization
of the Omega function and $k$--Efficient Power) in which
the Compromise Function has been used to select each one of their
generalization parameters (as it is showed in the appendices), just
like was done in the case of the generalized ecological function.
With this procedure we obtained three different objective functions
which lead the thermal engine at the same trade off operation mode.
Then these objective functions can be considered as particular cases
of the Compromise Function, with the advantage that we do not need
a second optimization process to reach the trade off optimal mode,
consequently an algebraic simplification is obtained.

\appendix

\section{NEHL Model}
\begin{figure}[t]
\begin{centering}
\includegraphics[width=15cm, height=6cm]{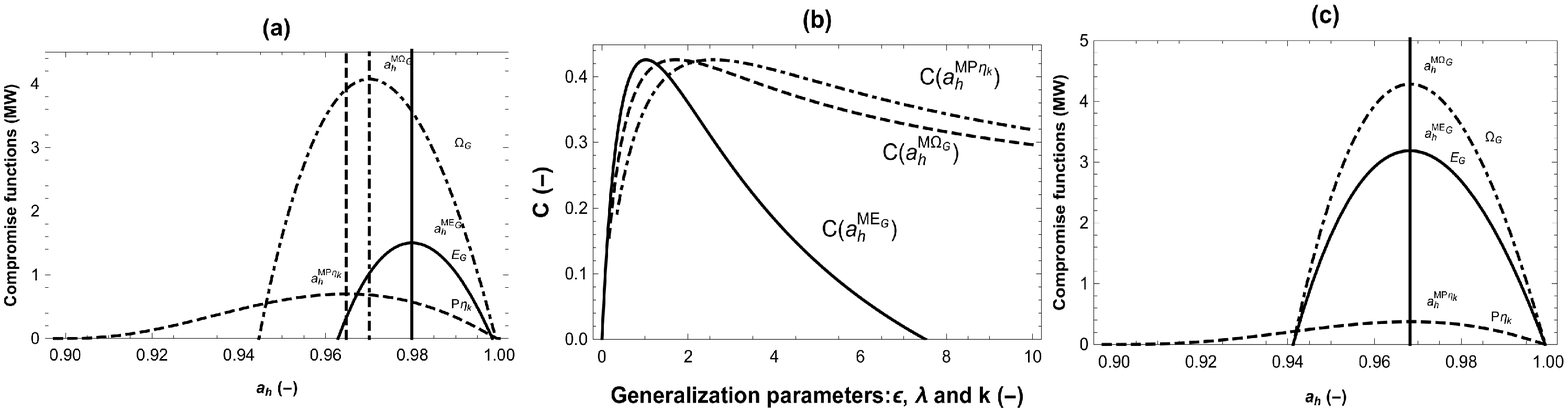}
\par\end{centering}
\caption{\label{fig:GOFNEHL}(a) Generalization of the objective functions
at the same value of their generalization parameters ($\epsilon=\lambda=k=2)$.
(b) Compromise function as a function of the generalization parameters
of the different generalization of the objective functions and (c)
Generalization of the compromise functions evaluated in the specific
values that maximize the compromise function ($\epsilon_{NEHL}^{MC_{P\Phi}}$,
$\lambda_{NEHL}^{MC_{P\Phi}}$ and $k_{NEHL}^{MC_{P\Phi}}$). Here
we use: $\alpha=1\,MW/K$, $\gamma=3$,$\tau=0.5$, $T_{1}=500\,K$,
$\delta=0.001\,MW/K$ and $R=0.9$}
\end{figure}

In section 2, we refer to certain calculus which were done in \cite{kPotEfi}.
They are related with the way to select each of the generalization
parameters by using the compromise function as is defined in equation
(\ref{eq:DefFunCompromiso}). When we use the generalizations of the
Ecological Function (equation \ref{eq:FunEcoGen}), the Omega Function
(equation \ref{eq:DefFunOmg}) or the $k$--Efficient Power
(equation \ref{eq:DefkPotEfi}), we get a family of each of these
functions associated to a generalization parameter respectively, to
pick up one of these parameters we use $C_{P\Phi}$. The functional
form of each one of these generalizations, can be obtained by considering
the equations (\ref{eq:PotNendoCCR}), (\ref{eq:DisMNoEndoRCC}) and
the efficiency, which is given by:
\begin{equation}
\eta\left(\alpha,\delta,\gamma,\tau,R,a_{h}\right)=\frac{\alpha\left(a_{h}-1\right)\left[a_{h}\left(R+\gamma\right)-\left(\gamma+\tau\right)\right]}{\left[a_{h}\left(R+\gamma\right)-\gamma\right]\left[\alpha\left(a_{h}-1\right)+\delta\left(\tau-1\right)\right]}.\label{eq:EfiMNoendoRCC}
\end{equation}
In this model, the power output is not the only characteristic function
which has a $a_{h}^{MP}$, the efficiency has a high reduced temperature
that maximize it and is:
\begin{equation}
a_{h}^{M\eta}\left(\alpha,\delta,\gamma,\tau,R\right)=\frac{\kappa\left[\gamma\delta\left(\tau-1\right)+\alpha\tau\right]-\sqrt{R\kappa\delta\left(\tau-1\right)\tau\left\{ \gamma\delta\left(\tau-1\right)+\alpha\tau-R\left[\alpha+\delta\left(1-\tau\right)\right]\right\} }}{\kappa\left[\kappa\delta\left(\tau-1\right)+\alpha\tau\right]}.\label{eq:ahMaxEfiMNoEndoRCC}
\end{equation}
This allow us to know the functional form to the generalizations of
the objective functions which were mentioned through this paper. In
this model, the generalization of the ecological function is:
\begin{equation}
E_{G}\left(\alpha,\delta,\gamma,T_{1},\tau,\epsilon,R,ah\right)=\frac{T_{1}\left(d_{2EG}a_{h}^{2}-d_{1EG}a_{h}+d_{0EG}\right)}{\gamma-a_{h}\left(R+\gamma\right)},\label{eq:EcoGenEpsNendoRCC}
\end{equation}
where
\begin{equation}
d_{2EG}\left(\alpha,\epsilon,\gamma,R\right)=\alpha\left(R+\gamma\right)\left(1+\epsilon\tau\right),\label{eq:d2EGEcoGenNendoRCC}
\end{equation}
\begin{equation}
d_{1EG}\left(\alpha,\delta,\epsilon,\gamma,R\right)=R\left[\alpha\left(1+\epsilon\tau\right)-\delta\epsilon\left(\tau-1\right)^{2}\right]+\alpha\left[2\gamma\left(1+\epsilon\tau\right)+\tau\left(1+\epsilon\right)\right]-\gamma\delta\epsilon\left(\tau-1\right)^{2},\label{eq:d1EGEcoGenNendoRCC}
\end{equation}
\begin{equation}
d_{0EG}\left(\alpha,\delta,\epsilon,\gamma,R\right)=-\gamma\delta\epsilon\left(\tau-1\right)^{2}+\alpha\left[\gamma\left(1+\epsilon\tau\right)+\tau\left(1+\epsilon\right)\right].\label{eq:d0EGEcoGenNendoRCC}
\end{equation}
As is possible to observe in Fig. \ref{fig:GOFNEHL} (a), this function
has a $a_{h}$which maximize it, and is given by:
\begin{equation}
a_{h}^{ME_{G}}\left(\gamma,\tau,\epsilon,R\right)=\frac{\gamma\left(1+\epsilon\tau\right)+\sqrt{R\tau\left(1+\epsilon\right)\left(1+\epsilon\tau\right)}}{\left(R+\gamma\right)\left(1+\epsilon\tau\right)}.\label{eq:ahMaxFunEcoGenEpsNendoRCC}
\end{equation}
On the other hand, to know the functional form to the generalization
of the omega function, we need to use the equations: (\ref{eq:QEnEndoCCR}),
(\ref{eq:PotNendoCCR}), (\ref{eq:EfiMNoendoRCC}) and (\ref{eq:ahMaxEfiMNoEndoRCC}),
because the $E_{u,eff}=E_{u}-z_{min}E_{i}$ and $E_{u,l}=z_{max}E_{i}-E_{u}$,
where $E_{u}$is the useful energy (in this case the Power Output),
$E_{i}$fucntion is the energy which is get in to the system ($Q_{i}$),
$z_{min}$ is the minimum of the efficiency, that in the case of the
heat engines is zero, and $z_{max}$is the efficiency (equation (\ref{eq:EfiMNoendoRCC})
) evaluated in the $a_{h}^{M\eta}$ (equation (\ref{eq:ahMaxEfiMNoEndoRCC}))
\cite{FunOmg}. Then $E_{u,eff}$ and $E_{u,l}$ are:
\begin{equation}
E_{u,eff}\left(\alpha,\gamma,T_{1},\tau,R,a_{h}\right)=\frac{T_{1}\alpha\left(a_{h}-1\right)\left[\left(\gamma+\tau\right)-a_{h}\kappa\right]}{a_{h}\kappa-\gamma},\label{eq:EneUtilNendoRCC}
\end{equation}

\begin{equation}
E_{u,l}\left(\alpha,\delta,\gamma,T_{1},\tau,R,a_{h}\right)=T_{1}\alpha\left[\left(a_{h}-1\right)\left(1-\frac{\tau}{\kappa a_{h}-\gamma}\right)-\frac{n_{e,p}}{d_{e,p}}\right],\label{eq:EneUtilPerNendoRCC}
\end{equation}
whit $\kappa=R+\gamma$ and :
\begin{equation}
r_{u,p}\left(\alpha,\delta,\gamma,\tau,R\right)=\sqrt{R\kappa\delta\tau\left(\tau-1\right)\left[\gamma\delta\left(\tau-1\right)+\alpha\tau-R\left(\alpha+\delta-\delta\tau\right)\right]},\label{eq:DefrEneUtilPerdidaNendoRCC}
\end{equation}

\begin{equation}
n_{e,p}\left(\alpha,\delta,\gamma,\tau,R\right)=\left\{ \begin{array}{c}
\left[\left(a_{h}-1\right)\alpha+\delta\left(\tau-1\right)\right]\\
\times\left[R\delta\kappa\left(\tau-1\right)+r_{u,p}\right]\\
\times\left\{ \tau\left[\tau\left(\delta\kappa+\alpha\right)-R\left(\alpha+\delta\right)-\delta\gamma\right]+r_{u,p}\right\} 
\end{array}\right\} ,\label{eq:NumEneUtilPerdidaNendoRCC}
\end{equation}

\begin{equation}
d_{e,u}\left(\alpha,\delta,\gamma,\tau,R\right)=\left(r_{u,p}-R\alpha\tau\right)\left\{ \begin{array}{c}
R\delta\left[\alpha\left(\gamma-\tau\right)-2\gamma\delta\left(\tau-1\right)\right]\left(\tau-1\right)\\
-\gamma\delta\left(\tau-1\right)\left[\gamma\delta\left(\tau-1\right)+\alpha\tau\right]\\
+R^{2}\delta\left(\tau-1\right)\left(\alpha+\delta-\delta\tau\right)+r_{u,p}
\end{array}\right\} .\label{eq:DenEneUtilPerdidaNendoRCC}
\end{equation}
Substituting equations (\ref{eq:EneUtilNendoRCC}) and (\ref{eq:EneUtilPerNendoRCC})
in the equation (\ref{eq:DefFunOmg}), it is possible to know the
functional form of the omega function. In Figure \ref{fig:GOFNEHL}
(a), we can appreciate that this function has a high reduced temperature
which maximize it given by:
\begin{equation}
a_{h}^{M\Omega_{G}}\left(\alpha,\delta,\gamma,\tau,R\right)=\frac{\gamma\kappa d_{\Omega G}+r_{\Omega G}}{\kappa^{2}d_{\Omega G2}},\label{eq:ahMaxFunOmgGendNendoRCC}
\end{equation}
where :
\begin{equation}
d_{\Omega G}\left(\alpha,\delta,\gamma,\tau,R\right)=\left\{ \begin{array}{c}
R^{2}A_{22}\left(\alpha+\delta-\delta\tau\right)+\gamma\delta\left(\tau-1\right)\left[\gamma\delta\left(1+\lambda\right)\left(\tau-1\right)+\alpha\lambda\tau\right]\\
+2\alpha\lambda r_{up}+R\left[A_{12}+\alpha^{2}\tau\lambda-\alpha\delta\gamma\left(\tau-1\right)\left(1+\lambda\right)\right]
\end{array}\right\} ,\label{eq:d2OmgGenNendoRCC}
\end{equation}

\begin{equation}
r_{\Omega G}\left(\alpha,\delta,\gamma,\tau,R\right)=\sqrt{R\kappa^{2}\tau\left(1+\lambda\right)\left[\gamma\delta+R\left(\alpha+\delta\right)-\kappa\delta\tau\right]^{2}d_{\Omega G2}},\label{eq:r1OmgGenNendoRCC}
\end{equation}

\begin{equation}
A_{22}\left(\alpha,\delta,\gamma,\tau,R\right)=\alpha-\delta\left(1+\lambda\right)\left(\tau-1\right),\label{eq:A22OmgGNendoRCC}
\end{equation}

\begin{equation}
A_{12}\left(\alpha,\delta,\gamma,\tau,R\right)=\alpha^{2}\tau+\delta^{2}\left[\tau-1\right]^{2}\left[2\gamma\left(2+\lambda\right)+\tau\right]+\alpha\delta\left(\tau-1\right)\left(2\lambda\tau-3\gamma\right)\label{eq:A12OmgGNendoRCC}
\end{equation}
and $r_{u,p}$ is given by the equation (\ref{eq:DefrEneUtilPerdidaNendoRCC}).

The $k$--Efficient Power, can be obtained by substituting
the equations (\ref{eq:PotNendoCCR}) and (\ref{eq:EfiMNoendoRCC})
in (\ref{eq:DefkPotEfi}), and we get:
\begin{equation}
P\eta_{k}\left(\alpha,\delta,\gamma,T_{1},\tau,R,k,a_{h}\right)=T_{1}\left[\alpha\left(1-a_{h}\right)+\delta\left(1-\tau\right)\right]\left\{ \frac{\alpha\left(a_{h}-1\right)\left[a_{h}\left(R+\gamma\right)-\left(\gamma+\tau\right)\right]}{\left[a_{h}\left(R+\gamma\right)-\gamma\right]\left[\delta\left(\tau-1\right)+\alpha\left(a_{h}-1\right)\right]}\right\} ^{1+k}.\label{eq:KPotEfiNEHL}
\end{equation}
As in the above generalizations, this function also has a $a_{h}$
that maximizes it (see Fig. \ref{fig:GOFNEHL} (a)), it is given by:
\begin{equation}
a_{h}^{MP\eta_{k}}\left(\alpha,\delta,\gamma,\tau,R,k\right)=\frac{2}{3}\sqrt{b_{2}^{2}-3d_{1}}\cos\left(\xi_{k}\right)-\frac{d_{2}}{3},\label{eq:ahmaxkpeNEHL}
\end{equation}
where $\xi_{k}$ is:
\[
\xi_{k}\left(\alpha,\delta,\gamma,\tau,R,k\right)=\frac{1}{3}\left\{ \pi+\arccos\left[\frac{6d_{2}^{3}-27\left(d_{2}d_{1}\right)+81d_{0}}{6\left(d_{2}^{2}-3d_{1}\right)^{3/2}}\right]\right\} ,
\]
and the coefficients $d_{2}$, $d_{1}$ and $d_{0}$ are:
\begin{equation}
d_{2}\left(\alpha,\delta,\gamma,\tau,R,k\right)=\frac{\alpha\left(k\tau-3\gamma\right)+R\delta\tau\left(1+k\right)+\gamma\delta\left(1+k\right)\left(\tau-1\right)-R\left[\alpha+\delta\left(1+k\right)\right]}{\alpha\left(R+\gamma\right)},\label{eq:Coefd2ParaAhMaxkPotEfiNendoRCC}
\end{equation}

\begin{equation}
d_{1}\left(\alpha,\delta,\gamma,\tau,R,k\right)=\frac{\alpha\gamma\left(2R+3\gamma\right)+2\gamma\delta\kappa\left(1+k\right)-\text{\ensuremath{\tau\left[\alpha\left(R+2k\kappa\right)+2\gamma\delta\kappa\left(1+k\right)\right]}}}{\alpha\left(R+\gamma\right)^{\text{2}}}\label{eq:Coefd1ParaAhMaxjPotEfiNendoRCC}
\end{equation}
and
\begin{equation}
d_{o}\left(\alpha,\delta,\gamma,\tau,R,k\right)=\frac{\alpha\tau\left(R+k\kappa\right)+\delta\left(\tau-1\right)\left(\gamma^{2}-R\tau\right)-\alpha\gamma^{2}}{\alpha\left(R+\gamma\right)^{\text{2}}}.\label{eq:Coefd0ParaAhMaxkPotEfiNendoRCC}
\end{equation}

In Fig. \ref{fig:GOFNEHL} (a),
each of the generalization is sketched
to the same value of their generalization parameters, however, no
one of them have their maximums in the same place, so, their optimal
modes of operations are not the same. To chose a value of each of
the generalized parameters, in 2001 \cite{FunCom2}, 2006 \cite{OmegaGen}
and 2016 \cite{kPotEfi} the compromise function (equation (\ref{eq:DefFunCompromiso}))
was used. This was possible when replacing the values of $a_{h}$
which maximize each of the generalizations as was done for the $ME_{G}$--regime. It is possible to appreciate in Fig.
\ref{fig:GOFNEHL} (b), that are values of $\epsilon$, $\lambda$
and $k$ which maximize the compromise function in each case, and
they are:
\begin{equation}
\epsilon_{NEHL}^{MC}=\frac{\alpha\left(R+\tau-2\sqrt{R\tau}\right)}{\gamma\delta\left(\tau-1\right)^{\text{2}}-\alpha\left[\tau+\sqrt{R\tau}\left(\tau-1\right)\right]+R\left[\delta\left(\tau-1\right)^{2}-\alpha\tau\right]},\label{eq:EpsMaxFunCompPotDisNendoRCC}
\end{equation}

\begin{equation}
\lambda_{NEHL}^{MC}=\frac{\left[2\sqrt{R\tau}-\left(R+\tau\right)\right]\left\{ \gamma\delta\left(\tau-1\right)-R\left[\alpha-\delta\left(\tau-1\right)\right]^{2}\right\} }{\left\{ 2\left[r_{up}+R\alpha\tau\right]\left[R\left(\alpha+\delta\right)+\delta\left(\gamma-\kappa\tau\right)\right]\right\} d_{\lambda c}\sqrt{R\tau}},\label{eq:LambdaMaxComPotDIsNendoRCC}
\end{equation}
and
\begin{equation}
k_{NEHL}^{MC}=\frac{\left\{ \delta\left[R^{2}-\gamma\right]\left[1-\tau\right]\left[2\sqrt{R\tau}-\tau\right]+\alpha\tau\left[3\sqrt{R\tau}-\tau\right]+R\left[\delta\left(1-\tau\right)\left(\gamma+\tau-2\sqrt{R\tau}\right)+\alpha\left(\sqrt{R\tau}-3\tau\right)\right]\right\} n_{k1}}{\left\{ \alpha\left[\tau-\sqrt{R\tau}\right]-\kappa\delta\left[1-\tau\right]\right\} \left\{ \sqrt{R\tau}\left[\alpha\tau-\gamma\delta\left(1-\tau\right)\right]\left[\alpha\tau-2\gamma\delta\left(1-\tau\right)\right]+d_{k1}-d_{k2}\right\} }\label{eq:kmaxComPotDisNendoRCC}
\end{equation}
with $r_{Cp}$ given by the equation (\ref{eq:rCpNendoRCC}) and :
\begin{equation}
d_{\lambda c}\left(\alpha,\delta,\gamma,\tau,R,k\right)=\left\{ \begin{array}{c}
2\alpha r_{up}+\gamma\delta\left(\tau-1\right)\left[2\gamma\text{\ensuremath{\delta\left(\tau-1\right)}+\ensuremath{\alpha\tau}}\right]\\
+R\left[\alpha^{2}\tau+4\gamma\delta^{2}\left(\tau-1\right)^{2}-\alpha\delta\left(3\gamma-\tau\right)\left(\tau-1\right)\right]+R^{2}\left[\alpha-2\delta\left(\tau-1\right)\right]
\end{array}\right\} ,\label{eq:dLamCNedoRCC}
\end{equation}

\begin{equation}
n_{k1}\left(\alpha,\delta,\gamma,\tau,R,k\right)=\alpha r_{Cp}+\left[\alpha\left(\sqrt{R\tau}-\tau\right)-\gamma\delta\left(1-\tau\right)\right]\left[\delta\kappa\left(1-\tau\right)+\alpha\left(\sqrt{R\tau}-\tau\right)\right],\label{eq:nk1NEHL}
\end{equation}

\begin{equation}
d_{k1}\left(\alpha,\delta,\gamma,\tau,R,k\right)=R^{2}\left[\alpha+2\delta\left(1-\tau\right)\right]\left[\alpha\tau+\delta\sqrt{R\tau}\left(1-\tau\right)\right]+2r_{Cp}\left[\alpha\left(\tau-\sqrt{R\tau}\right)-\gamma\delta\left(1-\tau\right)\right]\label{eq:dk1NEHL}
\end{equation}
and
\begin{equation}
d_{k2}\left(\alpha,\delta,\gamma,\tau,R,k\right)=R\left[2r_{Cp}\delta\left(1-\tau\right)-4\gamma\delta^{2}\left(1-\tau\right)^{2}\sqrt{R\tau}+\alpha^{2}\tau\left(\tau+\sqrt{R\tau}\right)+\alpha\delta\left(1-\tau\right)\left(3\tau\sqrt{R\tau}-\gamma\left[2\tau+\sqrt{R\tau}\right]\right)\right].\label{eq:dk2NEHL}
\end{equation}
When these values are substituted in their corresponding $a_{h}$,
the high reduced temperature which arise is the same:
\begin{equation}
a_{h}^{NEHL}=\frac{\gamma}{R+\gamma}-\frac{r_{Cp}-\sqrt{R\tau\left\{ \alpha\sqrt{R\tau}-\gamma\delta\left(1-\tau\right)-R\left[\alpha+\delta\left(1-\tau\right)\right]\right\} \left[\alpha\left(\tau-\sqrt{R\tau}\right)-\kappa\delta\left(1-\tau\right)\right]}}{\left[R+\gamma\right]\left[\alpha\left(\tau-\sqrt{R\tau}\right)-\kappa\delta\left(1-\tau\right)\right]}.\label{eq:ahMFoGNEHL}
\end{equation}
This can be seen in the Fig. \ref{fig:GOFNEHL} (c) where the three
objective functions are evaluated in their corresponding values $\epsilon_{NEHL}^{MC_{P\Phi}}$, $\lambda_{NEHL}^{MC_{P\Phi}}$
and $k_{NEHL}^{MC_{P\Phi}}$ which maximize the compromise function.

\section{IHL Model}
As in the above section, we will show how, using the compromise function,
as is defined in equation (\ref{eq:DefFunCompromiso}) with the k--Efficient
Power, the generalization of the ecological function and the generalization
of the Omega Function, we can get the same high reduced temperature
given by equation (\ref{eq:ahmCpPhi4C}) obtained in section 3, where
the compromise function is used directly as an objective function.

\begin{figure}[t]
\begin{centering}
\includegraphics[width=15cm, height=6cm]{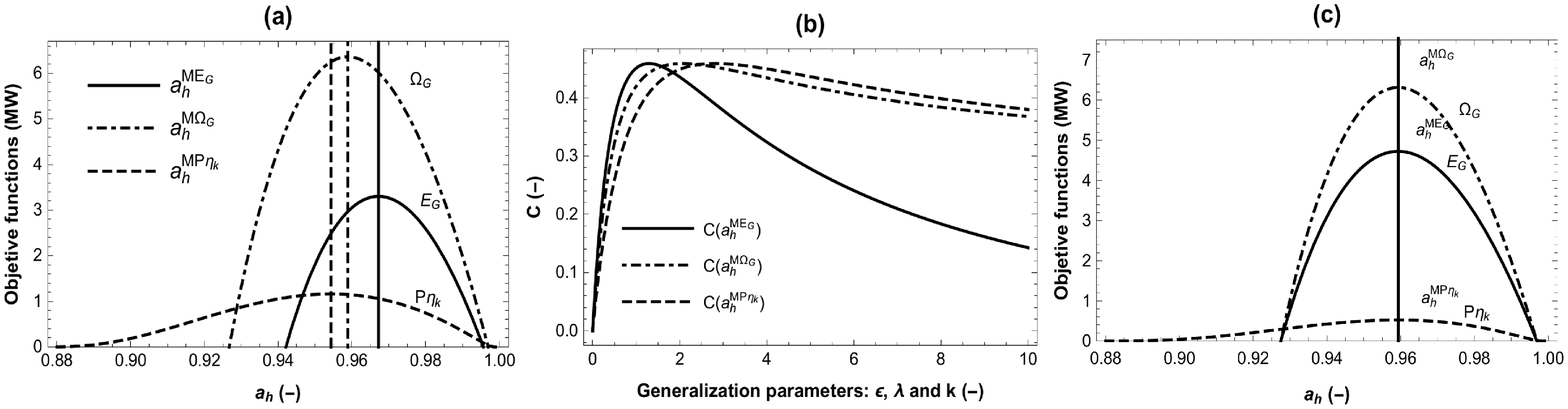}
\par\end{centering}
\caption{\label{fig:GOFIHL}(a) Generalization of the objective functions at
the same value of their generalization parameters ($\epsilon=\lambda=k=2)$.
(b) Compromise function as function of the generalization parameters
of the different generalization of the objective functions and (c)
Generalization of the compromise functions evaluated in the specific
value that maximize the compromise function ($\epsilon_{IHL}^{MC_{P\Phi}}$,
$\lambda_{IHL}^{MC_{P\Phi}}$ and $k_{IHL}^{MC_{P\Phi}}$). Here
we use:: $\alpha=1MW/K$, $\gamma=3$,$\tau=0.5$, $T_{1}=500K$,
$\delta=0.001MW/K$ and $r=0.001$}
\end{figure}

In this model, the functional form of efficiency is:
\begin{equation}
\eta\left(\alpha,\delta,\gamma,\tau,r,a_{h}\right)=\frac{\alpha\left(1-a_{h}\right)\left\{ \gamma+a_{h}\left[\gamma\left(r-1\right)-1\right]\right\} -\alpha\tau\left[1+a_{h}\left(r-1\right)\right]}{\left\{ \gamma+a_{h}\left[\gamma\left(r-1\right)-1\right]\right\} \left[\alpha\left(1-a_{h}\right)+\delta\left(1-\tau\right)\right]}.\label{eq:EfiMIrevCC}
\end{equation}
This efficiency has an $a_{h}$which maximize it, this is:
\begin{equation}
a_{h}^{M\eta}\left(\alpha,\delta,\gamma,\tau,r\right)=\frac{\alpha\tau+\gamma\left\{ \alpha\tau\left(1-r\right)+\delta\left(1-\text{\ensuremath{\tau}}\right)\left[\gamma\left(r-1\right)-1\right]\right\} -r_{\eta}}{\left[\gamma\left(r-1\right)-1\right]\left\{ \delta\left(\tau-1\right)\left[\gamma\left(r-1\right)-1\right]+\alpha\tau\left(r-1\right)\right\} },\label{eq:ahMaxEnfiIrevCC}
\end{equation}
where $r_{\eta}$ es:
\begin{equation}
r_{\eta}\left(\alpha,\delta,\gamma,\tau,r\right)=\sqrt{\tau\left[1-\gamma\left(r-1\right)\right]\left\{ r\tau\alpha^{2}+\delta^{2}\left(1-\tau\right)^{2}\left[1+\gamma\left(1-r\right)\right]-\alpha\delta\left(1-\tau\right)\left[\tau-1+r\left(\gamma-\tau\right)\right]\right\} }.\label{eq:rmetaIrevCC}
\end{equation}
This allow us to know the function form of the k--Efficient
Power, whicbh is:
\begin{equation}
P\eta_{k}\left(\alpha,\delta,\gamma,T_{1},\tau,r,a_{h}\right)=T_{1}\alpha^{k+1}\left\{ 1-a_{h}+\frac{\tau\left[1+a_{h}\left(r-1\right)\right]}{\gamma+a_{h}\left[\gamma\left(r-1\right)-1\right]}\right\} \left\{ \frac{\left(1-a_{h}\right)\left\{ \gamma+a_{h}\left[\gamma\left(r-1\right)-1\right]\right\} -\tau\left[1+a_{h}\left(r-1\right)\right]}{\left\{ \gamma+a_{h}\left[\gamma\left(r-1\right)-1\right]\right\} \left[\alpha\left(1-a_{h}\right)+\delta\left(1-\tau\right)\right]}\right\} ^{k}.\label{eq:kEfiPowIHL}
\end{equation}
In Fig. \ref{fig:GOFIHL} (a) this function is sketched to an arbitrary
value of $k$, and it is possible to appreciate that this function
has a high reduced temperature which maximize it, which is given by:
\begin{equation}
a_{h}^{MP\eta_{k}}\left(\alpha,\delta,\gamma,\tau,r\right)=\frac{2}{3}\sqrt{d_{2}^{2}-3d_{1}}\cos\left(\xi_{k}\right)-\frac{d_{2}}{3},\label{eq:ahmKPotEfiIHL}
\end{equation}
where $\xi_{k}$ is:
\begin{equation}
\xi_{k,}\left(\alpha,\delta,\gamma,\tau,r\right)=\frac{1}{3}\left\{ \pi+\arccos\left[\frac{6d_{2}^{3}-27\left(d_{2}d_{1}\right)+81d_{0}}{6\left(d_{2}^{2}-3d_{1}\right)^{3/2}}\right]\right\} ,\label{eq:anglemaxkPotEffiIHL}
\end{equation}
and the coefficients$d_{2}$, $d_{1}$ and $d_{0}$ are :
\begin{equation}
d_{2}\left(\alpha,\delta,\gamma,\tau,r\right)=-\frac{\delta\left(1-\tau\right)\left(1+k\right)\left[1+\gamma\left(1-r\right)\right]+\alpha\left[1+\gamma\left(3-r\right)-k\tau\left(1-r\right)\right]}{\alpha\left[1+\gamma\left(1-r\right)\right]},\label{eq:Coefd2ParaAhMaxkPotEfiICNC}
\end{equation}

\begin{equation}
d_{1}\left(\alpha,\delta,\gamma,\tau,r\right)=\frac{\alpha\gamma\left[2+\gamma\left(3-2r\right)\right]+2\gamma\delta\left(1+k\right)\left(1-\tau\right)\left[1+\gamma\left(1-r\right)\right]-\alpha\tau\left\{ 1+2k\gamma\left[1+\gamma\left(1-r\right)\right]\right\} }{\alpha\left[1+\gamma\left(1-r\right)\right]^{2}},\label{eq:Coefd1ParaAhMaxjPotEfiICNC}
\end{equation}
and 
\begin{equation}
d_{o}\left(\alpha,\delta,\gamma,\tau,r\right)=\frac{\alpha\left[\tau k\left(1+\gamma\right)+\tau-\gamma^{2}\right]-\delta\left(1+k\right)\left(\gamma^{2}-\tau\right)\left(1-\tau\right)}{\alpha\left[1+\gamma\left(1-r\right)\right]^{2}}.\label{eq:Coefd0ParaAhMaxkPotEfiICNC}
\end{equation}

For the generalization of the ecological function, it is necessary
to replace the equations (\ref{eq:Pot4C}) and (\ref{eq:Dis4C}) in
(\ref{eq:FunEcoGen}), having so, to this model $E_{G}$ is:
\begin{equation}
E_{G}\left(\alpha,\delta,\gamma,\tau,r,\epsilon\right)=T_{1}\frac{a_{h}^{2}\alpha\left[1+\epsilon\tau\right]\left[1+\gamma\left(1-r\right)\right]-a_{h}n_{EG1}-\gamma\delta\left[1-\tau\right]^{2}+\alpha\left[\gamma\left(1+\epsilon\tau\right)+\tau\left(1+\epsilon\right)\right]}{\gamma-a_{h}\left[1+\gamma\left(1-r\right)\right]},\label{eq:GenFunEcoIHL}
\end{equation}
with:
\[
n_{EG1}\left(\alpha,\delta,\gamma,\tau,r,\epsilon\right)=\alpha\left[1+\gamma\left(2-r\right)+\tau-\tau\left(r-\epsilon\left[1+\gamma\right]\left[2-r\right]\right)\right]-\delta\epsilon\left[1-\tau\right]^{2}\left[1+\gamma\left(1-r\right)\right]
\]
In Fig. \ref{fig:GOFIHL} (a), it is possible to appreciate that
to an arbitrary value of $\epsilon$, this function has a $a_{h}$which
maximize it, that is:
\begin{equation}
a_{h}^{ME_{G}}\left(\gamma,\tau,r,\epsilon\right)=\frac{\gamma\left(1-\epsilon\tau\right)+\sqrt{\tau\left(1+\epsilon\right)\left(1+\epsilon\tau\right)}}{\left(1+\epsilon\tau\right)\left[1+\gamma\left(1-r\right)\right]}.\label{eq:ahMaxGenEcoIHL}
\end{equation}
On the other hand, by taking the definitions of $E_{u,eff}$, $E_{u,l}$
and z \cite{FunOmg} for this model, we get:
\begin{equation}
E_{u,eff}\left(\alpha,\gamma,T_{1},\tau,r,a_{h}\right)=T_{1}\alpha\left\{ 1-a_{h}+\frac{\tau\left[1-a_{h}\left(1-r\right)\right]}{\gamma-a_{h}\left[1+\gamma\left(1-r\right)\right]}\right\} \label{eq:EneUtilIrrevCC}
\end{equation}
and
\begin{equation}
E_{u,l}\left(\alpha,\delta,\gamma,T_{1},\tau,r,a_{h}\right)=T_{1}\alpha\left\{ a_{h}-1+\frac{\tau\left[a_{h}\left(1-r\right)-1\right]}{\gamma-a_{h}\left[\gamma\left(1-r\right)+1\right]}+\frac{\left[\alpha\left(1-a_{h}\right)+\delta\left(1-\tau\right)\right]\left(r_{\eta}^{2}-r_{\eta}n_{1ep}-n_{2ep}\right)}{\left(r_{\eta}-\alpha\tau\right)\left\{ r_{\eta}\alpha-\left[\gamma\left(r-1\right)+1\right]d_{1ep}\right\} }\right\} ,\label{eq:EneUtilPerIrrevCC}
\end{equation}
where $r_{\eta}$ is given by the equation (\ref{eq:rmetaIrevCC}) and 
\begin{equation}
n_{1ep}\left(\alpha,\delta,\gamma,\tau,r\right)=\alpha\tau\left\{ 1+r\left[1+\gamma\left(1-r\right)\right]-\tau\left(1-r\right)^{2}\right\} -\delta\left(1-\tau\right)\left[\gamma\left(1-r\right)+1\right]\left[r\gamma-\tau\left(1-r\right)-1\right],\label{eq:Num1EneUtilPerdidaIrrevCC}
\end{equation}

\begin{equation}
n_{2ep}\left(\alpha,\delta,\gamma,\tau,r\right)=\tau\left[\gamma\left(r-1\right)+1\right]\left\{ \delta^{2}\left(1-\tau\right)^{2}\left[1+\gamma\left(1-r\right)\right]+r\alpha^{2}\tau-\alpha\delta\left(1-\tau\right)\left[r\left(\gamma-\tau\right)-1+\tau\right]\right\} ,\label{eq:Num2EneUtilPerdidaIrrevCC}
\end{equation}
with
\begin{equation}
d_{1ep}\left(\alpha,\delta,\gamma,\tau,r\right)=r\alpha^{2}\tau-\alpha\delta\left(1-\tau\right)\left[r\gamma+\tau\left(1-r\right)-1\right]+\delta^{2}\left(1-\tau\right)^{2}\left[1+\gamma\left(1-r\right)\right].\label{eq:Den2EneUtilPerdidaIrrevCC}
\end{equation}
In Fig. \ref{fig:GOFIHL} (a) can be appreciated that the generalization of the Omega Function has a maximum, which is given by:
\begin{equation}
a_{h}^{M\Omega_{G}}\left(\alpha,\delta,\gamma,\tau,r,\lambda\right)=\frac{\Omega_{2}-\sqrt{\Omega_{2}+4\Omega_{1}\Omega_{3}}}{2\Omega_{1}},\label{eq:ahmaxOmgGenIHL}
\end{equation}
with:
\begin{equation}
\Omega_{1}\left(\alpha,\delta,\gamma,\tau,r,\lambda\right)=-d_{\Omega g1}^{2}\left\{ r_{\eta}^{2}\alpha+\alpha\tau d_{\Omega g1}\left\{ \delta^{2}d_{\Omega g1}+r\tau\alpha^{2}+\left(1-\tau\right)\left[1-r\gamma-\tau\left(1-r\right)\right]\right\} +r_{\eta}n_{\Omega g1}\right\} ,\label{eq:CofOmgG1}
\end{equation}

\begin{equation}
\Omega_{2}\left(\alpha,\delta,\gamma,\tau,r,\lambda\right)=-\frac{2\gamma}{d_{\Omega g1}}\Omega_{1},\label{eq:CoefOmg2IHL}
\end{equation}

\begin{equation}
\Omega_{3}\left(\alpha,\delta,\gamma,\tau,r,\lambda\right)=\left\{ \begin{array}{c}
r_{\eta}^{2}\alpha\left[\tau\left(1+\lambda\right)-\gamma^{2}\right]-r_{\eta}d_{\Omega g2}\\
-\alpha\tau d_{\Omega g1}\left[\gamma^{2}-\tau\left(1+\lambda\right)\right]\left\{ \delta^{2}\left(1-\tau\right)^{2}d_{\Omega g1}+r\alpha^{2}\tau-\alpha\delta\left(1-\tau\right)\left[\tau-1+r\left(\gamma+\tau\right)\right]\right\} 
\end{array}\right\} ,\label{eq:CofOmgG3}
\end{equation}
and
\begin{equation}
d_{\Omega g1}\left(\gamma,r\right)=1+\gamma\left(1-r\right),\label{eq:dOmgG1}
\end{equation}

\begin{equation}
d_{\Omega g2}\left(\alpha,\delta,\gamma,\tau,r,\lambda\right)=\left\{
\begin{array}{c}
-\delta^{2}d_{\Omega g1}^{2}\left(\gamma^{2}-\tau\right)\left(1+\lambda\right)+\alpha\delta\left(1-\tau\right)d_{\Omega g1}n_{\Omega g2}\\
+\alpha^{2}\tau\left\{ -\gamma^{2}\left[1+r+r\gamma\left(1-r\right)\right]+\tau+\tau d_{\Omega g1}\left\{ r+\lambda\left[1+r-\gamma\left(1-r\right)\right]\right\} \right\} 
\end{array}
\right\} .\label{eq:dOmgG2}
\end{equation}

\[
n_{\Omega g1}\left(\alpha,\delta,\gamma,\tau,r,\lambda\right)=-\delta^{2}d_{\Omega g1}^{2}\left(1-\tau\right)^{2}-\alpha^{2}\tau\left[1+rd_{\Omega g1}+\lambda\tau\left(1-r\right)^{2}\right]+\alpha d_{\Omega g1}\left(1-\tau\right)\left[r\gamma-1+\tau\left(1-r\right)\left(1+2\lambda\right)\right]
\]
and
\[
n_{\Omega g2}\left(\gamma,\tau,r,\lambda\right)=r\gamma^{3}-r\gamma\tau\left(1+\lambda\right)+\tau\left(1+\lambda\right)\left[1-\tau\left(1-r\right)\right]-\gamma^{2}\left[1-\tau\left(1+2\lambda\right)\left(1-r\right)\right].
\]
When this high reduced temperatures ((\ref{eq:ahmKPotEfiIHL}), (\ref{eq:ahMaxGenEcoIHL})
and (\ref{eq:ahmaxOmgGenIHL})) are substituted in equation (\ref{eq:DefFunCompromiso}),
the compromise function in each one has a value of $k$, $\lambda$
and $\epsilon$ which maximize it, just like is showed in Fig. \ref{fig:GOFIHL}(b),
if the compromise function is maximized respect to each generalization
parameter. The corresponding values are:
\begin{equation}
\epsilon_{IHL}^{MC_{P\Phi}}=\frac{\alpha\left[\left(1-\sqrt{\tau}\right)^{2}-r\left(\gamma+\tau\right)\right]}{\delta\left[1+\gamma\left(1-r\right)\right]\left(1-\tau\right)^{2}+\alpha\tau^{1/2}\left[1+\tau^{1/2}\left(r+r\gamma-2\right)+\tau\right]}.\label{eq:EpsMaxFunCom4C}
\end{equation}

\begin{equation}
\lambda_{IHL}^{MC_{P\Phi}}=\frac{\left[\left(1-\sqrt{\tau}\right)^{2}-r\left(\gamma+\tau\right)\right]\left\{ 2\alpha\tau d_{\Omega g1}n_{c1}-r_{\eta}\left\{ \begin{array}{c}
\delta^{2}\left(1-\tau\right)^{2}d_{\Omega g1}+\alpha^{2}\tau\left[1+r+r\gamma\left(1-r\right)\right]\\
+\alpha\delta d_{\Omega g1}\left(1-\tau\right)\left[1-r\left(\gamma-\tau\right)+\tau\right]
\end{array}\right\} \right\} }{r_{\eta}\sqrt{\tau}d_{c1}-2\tau d_{\Omega g1}n_{c1}\left[\delta\left(1-\tau\right)d_{\Omega g1}+\alpha\sqrt{\tau}-\alpha\tau\left(1-r\right)\right]}\label{eq:LamMaxCPPhiIHL}
\end{equation}
and
\begin{equation}
k_{IHL}^{MC_{P\Phi}}=\frac{\left[\left(1-\sqrt{\tau}\right)^{2}-r\left(\gamma+\tau\right)\right]\left\{ \begin{array}{c}
\delta^{2}\left(1-\tau\right)^{2}d_{\Omega g1}^{2}+\alpha^{2}\left(1-r\gamma\right)\left[1-\sqrt{\tau}\left(1-r\right)\right]\\
+\alpha\delta d_{\Omega g1}\left[r\left(\gamma-\tau\right)-1-\sqrt{\tau}\left(1-\sqrt{\tau}\right)\right]
\end{array}\right\} -\alpha r_{kC}}{\left\{ \delta d_{\Omega g1}\left(1-\tau\right)+\alpha\left[\sqrt{\tau}-\tau\left(1-r\right)\right]\right\} \left\{ \sqrt{\tau}\left\{ \delta d_{\Omega g1}\left(1-\tau\right)-\alpha\left[\tau-1+r\left(\gamma-\tau\right)\right]\right\} -2r_{kC}\right\} }\label{eq:kMaxCpPhiIHL}
\end{equation}
with 
\begin{equation}
n_{c1}\left(\alpha,\delta,\gamma,\tau,r\right)=-\delta^{2}\left(1-\tau\right)^{2}d_{\Omega g1}-r\tau\alpha^{2}-\alpha\delta\left(1-\tau\right)\left[1-r\gamma-\tau\left(1-r\right)\right],\label{eq:nc1IHL}
\end{equation}

\begin{equation}
d_{c1}\left(\alpha,\delta,\gamma,\tau,r\right)=\left\{ \begin{array}{c}
-2\delta^{2}\left(1-\tau\right)^{2}d_{\Omega g1}^{2}-\alpha\delta d_{\Omega g1}\left[1+2\sqrt{\tau}+3r\tau-\left(r\gamma+3\tau\right)\right]\\
-\alpha^{2}\tau\left[1+r\left(1+\gamma\right)-r^{2}\gamma-2\sqrt{\tau}\left(1-r\right)+\tau\left(1-r\right)^{2}\right]
\end{array}\right\} \label{eq:dc1IHL}
\end{equation}
and
\begin{equation}
r_{kC}=\sqrt{\tau\left[\alpha\left(1-r\gamma+\sqrt{\tau}\right)+\delta\left(1-\tau d_{\Omega g1}\right)\right]\left[\delta\left(1-\tau d_{\Omega g1}\right)+\alpha\sqrt{\tau}-\alpha\tau\left(1-r\right)\right]}.\label{eq:rkcIHL}
\end{equation}

When these values of the generalization parameters are substituted
in the corresponding high reduced temperature that maximize the generalized
functions, it is possible to appreciate in Fig. \ref{fig:GOFIHL}(c),
that the three generalized functions have their maxima in the same
$a_{h}$, then the three $a_{h}$ reduces them self to:
\begin{equation}
a_{h}^{IHL}=\left[\frac{1}{1-\gamma\left(r-1\right)}\right]\left(\gamma+\sqrt{\frac{\tau\left\{ \delta\left(\tau-1\right)\left[\gamma\left(r-1\right)-1\right]-\alpha\left(r\gamma+\sqrt{\tau}-1\right)\right\} }{\delta\left(\tau-1\right)\left[\gamma\left(r-1\right)-1\right]+\alpha\left[\sqrt{\tau}+\tau\left(r-1\right)\right]}}\right).\label{eq:ahMaxCpPhiOldIhL}
\end{equation}
Which is just the same high reduced temperature that is get when the
compromise function is used directly like $P\Phi$--Compromise
Function (equation (\ref{eq:ahmCpPhi4C})), then they become equivalent
in their energetics at the optimal operation regimes.

\end{document}